\documentstyle[11pt,newpasp,epsf]{article}
\def\etal {{\it et~al.}}
\def\fd {\dot f}
\def\fdd {\ddot f}
\def\fddd {\stackrel {\ldots}{f}}
\def\fdddd {f^{(4)}}
\def\fddddd {f^{(5)}}

\begin{document}

\title{A Hierarchical Triple Star System in M4}

\author{Frederic A.~Rasio}

\affil{Department of Physics, MIT, Cambridge, MA 02139, USA}

\begin{abstract}
The radio millisecond pulsar PSR~B1620$-$26 is part of an extraordinary 
triple star system in the globular cluster M4. The inner companion to
the neutron star is thought to be a white dwarf of mass $m_1\simeq0.3\,M_\odot$
in an orbit of period $\simeq0.5\,$yr. The nature and
orbital characteristics of the second, more distant companion, have remained
a mystery for many years. A theoretical analysis of the 
latest available radio pulsar timing data is presented here, allowing us 
to determine approximately
the mass and orbital parameters of the second companion.
Remarkably, the current best-fit parameters correspond to a second companion of 
{\it planetary mass\/}, with
$m_2 \sin i_2 \simeq 7\times10^{-3}\,M_\odot$, in an orbit of eccentricity
$e_2\simeq 0.45$ and with a large semimajor axis $a_2\simeq 60\,$AU. 
The short dynamical lifetime of this very wide triple in M4 suggests that
large numbers of such planets must be present in globular clusters. 
We also address the question of the anomalously
high eccentricity of the inner binary pulsar. While this eccentricity could have been 
induced during the same dynamical interaction that created the triple, 
we find that it could also naturally
arise from long-term secular perturbation effects in the triple,
combining the general relativistic precession of
the inner orbit with the Newtonian gravitational perturbation by the outer
planet. 
\end{abstract}

\section{Introduction}

PSR~B1620$-$26 is a unique millisecond radio pulsar. The pulsar is a member of
a hierarchical triple system located in or near the
core of the globular cluster M4. It is the only radio pulsar known in a triple
system, and the only triple system known in any globular cluster. The inner binary 
of the triple contains the $\simeq 1.4\,M_{\odot}$ neutron star with 
a $\simeq 0.3\,M_{\odot}$ white-dwarf
companion in a 191-day orbit (Lyne \etal\ 1988; McKenna \& Lyne 1988).
The triple nature of the system was
first proposed by Backer (1993) in order to explain the unusually high residual 
second and third pulse frequency derivatives left over after subtracting a standard 
Keplerian model for the pulsar binary.

The pulsar has now been timed for 12 years since its discovery
(Thorsett, Arzoumanian, \& Taylor 1993; Backer, Foster, \& Sallmen 1993;
Backer \& Thorsett 1995; Arzoumanian \etal\ 1996; Thorsett \etal~1999). 
These observations have not only confirmed the triple nature of the system, 
but they have also provided constraints on the mass and orbital 
parameters of the second companion. Earlier calculations using three pulse
frequency derivatives suggested that the mass of the second companion could
be anywhere between $\sim10^{-3}-1\,M_\odot$, with corresponding
orbital periods in the range $\sim 10^2-10^3\,$yr (Michel 1994; Rasio 1994;  
Sigurdsson 1995). More recent calculations using four frequency derivatives
and preliminary measurements of the orbital perturbations of the inner
binary further constrained the mass of the second companion,
and suggested that it is most likely a giant planet or a brown dwarf of mass 
$\sim0.01\,M_\odot$ at a distance $\sim 50\,$AU from the pulsar 
binary (Arzoumanian \etal\ 1996; Joshi \& Rasio 1997, hereafter JR97).
In \S2 below we summarize our results from the latest theoretical analysis 
of the pulsar timing data (Ford \etal~2000; hereafter FJRZ), including the most recent
observations of Thorsett \etal\ (1999; hereafter TACL). The data now include 
measurements of
five pulse frequency derivatives, as well as improved measurements and 
constraints on various orbital perturbation effects in the triple. 

Previous optical observations by Bailyn \etal\ (1994) and 
Shearer \etal\ (1996) using ground-based
images of M4 had identified a possible optical counterpart for the pulsar,
consistent with a $\sim0.5\,M_\odot$ main-sequence star, thus
contradicting the theoretical results, which suggest a much lower-mass
companion. However, it also seemed possible that the object could be a blend
of unresolved fainter stars, if not a chance superpositon. 
Later HST WFPC2 observations of the same region by Bailyn (private communication) 
have resolved the uncertainty. The much higher resolution ($\sim0.1\arcsec$) HST image
shows no optical counterpart at the pulsar position, down to a magnitude 
of ${\rm V}\simeq23$,
therefore eliminating the presence of any main-sequence star in the system. 

PSR~B1620$-$26 is not the first millisecond pulsar system in which a 
planet (or brown dwarf) has been detected. The first one, PSR~B1257+12, is a
single millisecond pulsar with three clearly detected inner planets 
(all within $1\,$AU) of terrestrial masses in circular orbits around the
neutron star (Wolszczan 1994). Preliminary evidence for
at least one giant planet orbiting at a much larger distance from the
neutron star has also been reported (Wolszczan
1996; JR97). 
In PSR~B1257+12 it is likely that the
planets were formed in orbit around the neutron star, perhaps out of
a disk of debris left behind following the complete evaporation of a
previous binary companion (see, e.g., Podsiadlowski 1995). 
Such an evaporation process has been observed in several eclipsing
binary millisecond pulsars (e.g., Nice \etal\ 2000),
where the companion masses have 
been reduced to $\sim0.01\,M_\odot$ by ablation. These companions
used to be ordinary white dwarfs and, although their masses are now
quite low, they cannot be properly called either planets or brown dwarfs
(Rasio, Pfahl, \& Rappaport 2000).

In PSR~B1620$-$26, the hierarchical triple configuration of the system and its
location near the core of a dense globular cluster suggest that the second
companion was acquired by the pulsar following a dynamical interaction with another
object (Rasio, McMillan \& Hut 1995; Sigurdsson 1993, 1995; FJRZ). 
This object could have been 
a primordial binary with a low-mass brown-dwarf component, or a main-sequence star
with a planetary system containing at least one massive giant planet.
Indeed the possibility of detecting ``scavenged'' planets around millisecond
pulsars in globular clusters was discussed by Sigurdsson (1992) even before
the triple nature of PSR~B1620$-$26 was discovered.
Several versions of such a dynamical formation scenario are possible for 
the triple system,
all involving dynamical exchange interactions between binaries in the
core of M4. In the most likely scenario, studied in detail by FJRZ,
a pre-existing binary millisecond pulsar has a dynamical interaction
with a wide star--planet system, which leaves the planet bound to the
binary pulsar while the star is ejected. From numerical
scattering experiments we found that the probability of retaining
the planet, although smaller than the probability of retaining the star,
is always significant, with a branching ratio $\simeq10\%-30\%$ for encounters
with pericenter distances $r_p/a_i \simeq 0.2-1$, 
where $a_i\sim50\,$AU is the typical initial star--planet separation.
All the observed parameters of the triple system are consistent with
such a formation scenario, which also allows the age of the millisecond pulsar
(most likely $\ga10^9\,$yr) to be much larger than the lifetime of the triple
(as short as $\sim 10^7\,$yr if it resides in the core of the cluster).

Objects with masses $\sim0.001-0.01\,M_\odot$ have recently been detected
around many nearby solar-like stars in Doppler searches for extrasolar planets
(see Marcy \& Butler 1998 for a review).
In several cases (e.g., $\upsilon$ And), more than one object have been detected
in the same system, clearly establishing that they are members of a planetary
system rather than a very low-mass stellar (brown dwarf) binary companion. 
For the second companion of PSR~B1620$-$26,
of mass $\sim0.01\,M_\odot$, current observations and theoretical modeling do
not make it possible to determine whether the object was originally formed as
part of a planetary system, or as a brown dwarf.
In this paper, we will simply follow our prejudice, and henceforth will refer to
the object as ``the planet.''

One aspect of the system that remains unexplained, and can perhaps provide
constraints on its formation and dynamical evolution, is the unusually high 
eccentricity $e_1 = 0.025$ of the inner binary. This is much larger than 
one would expect for a binary millisecond pulsar formed through the standard
process of pulsar recycling by accretion from a red-giant companion. During
the mass accretion phase, tidal circularization of the orbit through turbulent
viscous dissipation in the red-giant envelope should have brought the eccentricity
down to $\la10^{-4}$ (Phinney 1992). At the same time, however, the measured
value may appear too small for a dynamically induced eccentricity.
Indeed, for an initially circular binary, the eccentricity induced by 
a dynamical interaction with another star is an extremely steep function of the 
distance of closest approach (Rasio \& Heggie 1995). Therefore a ``typical'' interaction
would be expected either to leave the eccentricity unchanged, 
or to increase it to a value of
order unity (including the possibility of disrupting the binary). 

Secular perturbations in the triple system can also lead to an increase in the
eccentricity of the inner binary. A previous analysis assuming nearly coplanar orbits
suggested that, starting from a circular inner orbit, an eccentricity as large as 
0.025 could only be induced by the
perturbation from a stellar-mass second companion (Rasio 1994), which is now ruled
out. For large relative
inclinations, however, it is known that the eccentricity perturbations can in
principle be considerably larger (Kozai 1962; see Ford, Kozinsky \& Rasio 2000,
hereafter FKR, for a recent treatment). 
In addition, the Newtonian secular perturbations due to the tidal
field of the second companion can combine
nonlinearly with other perturbation effects, such as the general relativistic
precession of the inner orbit, to produce enhanced eccentricity perturbations
(see FKR and \S3 below).

\section{Analysis of the Pulsar Timing Data}

\subsection{Pulse Frequency Derivatives}

The standard method of fitting a Keplerian orbit to timing residuals
cannot be used when the pulsar timing data cover only a small fraction of
the orbital period (but see TACL for an attempt at fitting
two Keplerian orbits to the PSR~B1620$-$26 timing data). For PSR~B1620$-$26, 
the duration of the observations
($\simeq10\,$yr) is short compared to the likely orbital period of
the second companion, which is $\ga100\,$yr. In this case, 
pulse frequency derivatives (coefficients in a Taylor expansion of 
the pulse frequency around a reference epoch) can be derived to characterize the 
shape of the timing residuals (after subtraction of a Keplerian model for the inner
binary). It is easy to show that from {\it five\/} well-measured
and dynamically-induced frequency derivatives one can obtain a complete
solution for the orbital parameters and mass of the companion, up to the
usual inclination factor (JR97).

JR97 used the first 
four time derivatives of the pulse frequency to solve for a one-parameter family
of solutions for the orbital parameters and mass of the second companion.
The detection of the fourth derivative, 
which was marginal at the time, has now been confirmed (TACL). 
In addition, we now also have a preliminary
measurement of the fifth derivative. This allows us in principle
to obtain a 
unique solution, but the measurement uncertainty on the fifth derivative 
is very large, giving us correspondingly large uncertainties on the
theoretically derived parameters of the system. Equations and details on the method
of solution are presented in JR97, and will not be repeated here.

Our new solution is based on the latest available values of 
the pulse frequency derivatives, obtained by TACL 
for the epoch MJD 2448725.5:
\begin{eqnarray}
{\rm Spin\, Period}\, P & = & 11.0757509142025(18) \,{\rm ms} \nonumber \\
{\rm Spin\, frequency}\, f & = & 90.287332005426(14) \, {\rm s}^{-1} \nonumber \\
\fd & = & -5.4693(3)\times10^{-15}\,  {\rm s}^{-2} \nonumber \\
\fdd & = & 1.9283(14)\times10^{-23} \, {\rm s}^{-3} \nonumber \\
\fddd & = & 6.39(25)\times10^{-33} \, {\rm s}^{-4} \nonumber \\
\fdddd & = & -2.1(2)\times10^{-40} \, {\rm s}^{-5} \nonumber \\
\fddddd & = & 3(3)\times10^{-49} \, {\rm s}^{-6} \nonumber
\end{eqnarray}
Here the number in parenthesis is a conservative estimate of the formal $1\sigma$ error 
on the measured best-fit value, taking into account the correlations between
parameters (see TACL for details). It should be noted that the best-fit value 
for the fourth derivative quoted earlier by Arzoumanian \etal\
(1996) and used in JR97, $\fdddd = -2.1(6)\times10^{-40} \, {\rm s}^{-5}$,
has not changed, while the estimated $1\sigma$ error has decreased by a factor of three.
This gives us confidence that the new measurement of $\fddddd$, although 
preliminary, will not change significantly over the next few years as more
timing data become available. 

Since the orbital period of the second companion is much
longer than that of the inner binary, we treat the inner binary as a single object. 
Keeping the same notation as in JR97, we let $m_1 = m_{\rm NS} + m_c$ be the mass of the 
inner binary pulsar, with $m_{\rm NS}$ the mass of the neutron star and $m_c$ the mass 
of the (inner) companion, and we denote by $m_2$ the mass of the second companion.
The orbital parameters are the longitude $\lambda_2$ at epoch (measured from 
pericenter), the longitude of pericenter $\omega_2$ (measured from the ascending 
node), the eccentricity $e_2$, semimajor axis $a_2$, and inclination $i_2$ (such 
that $\sin i_2=1$ for an orbit seen edge-on). They all
refer to the orbit of the second companion with respect to the center of 
mass of the system (the entire triple). A subscript 1 for the 
orbital elements refers to the orbit of the inner binary.
We assume that $m_{\rm NS}=1.35\,M_\odot$, giving $m_c\sin i_1 =0.3\,M_\odot$, 
where $i_1$ is the inclination of the inner binary (Thorsett \etal\ 1993),
and we take $\sin i_1=1$ for the analysis presented in this section since 
our results depend only very weakly on the inner companion mass.

The observed value of $\fd$ is in general determined by a combination of the
intrinsic spin-down of the pulsar and the acceleration due to the second 
companion. However, in this case, the observed value of $\fd$ has changed
from $-8.1\times10^{-15}\rm{s^{-2}}$ to $-5.4\times10^{-15}\rm{s^{-2}}$
over 11 years (TACL). Since the intrinsic spin-down rate is essentially
constant,
this large observed rate of change indicates that the observed $\fd$ is 
almost entirely acceleration-induced. Similarly, the observed value of $\fdd$ 
is at least an order of magnitude larger than the estimate of $\fdd$
from intrinsic timing noise, which is usually not measurable
for old millisecond pulsars (see Arzoumanian \etal\ 1994, TACL, and JR97). 
Intrinsic contributions to the higher derivatives
should also be completely negligible for millisecond pulsars. 
Hence, in our analysis, we assume that all observed frequency derivatives 
are dynamically induced, reflecting the presence of the second companion.

Figure~1 illustrates our latest one-parameter family of solutions,
obtained using the updated values of the first 
four pulse frequency derivatives. There are no significant differences
compared to the solution obtained previously in JR97.
The vertical solid line indicates the unique solution obtained by including the 
fifth derivative. It corresponds to a second companion mass
$m_2 \sin i_2 = 7.0\times10^{-3}\,M_\odot$, eccentricity
$e_2=0.45$, and semimajor axis $a_2=57\,$AU. For a total system mass
$m_1+m_2=1.65\,M_\odot$ this gives an outer orbital period $P_2=308\,$yr.

\begin{figure}[t]
\plottwo{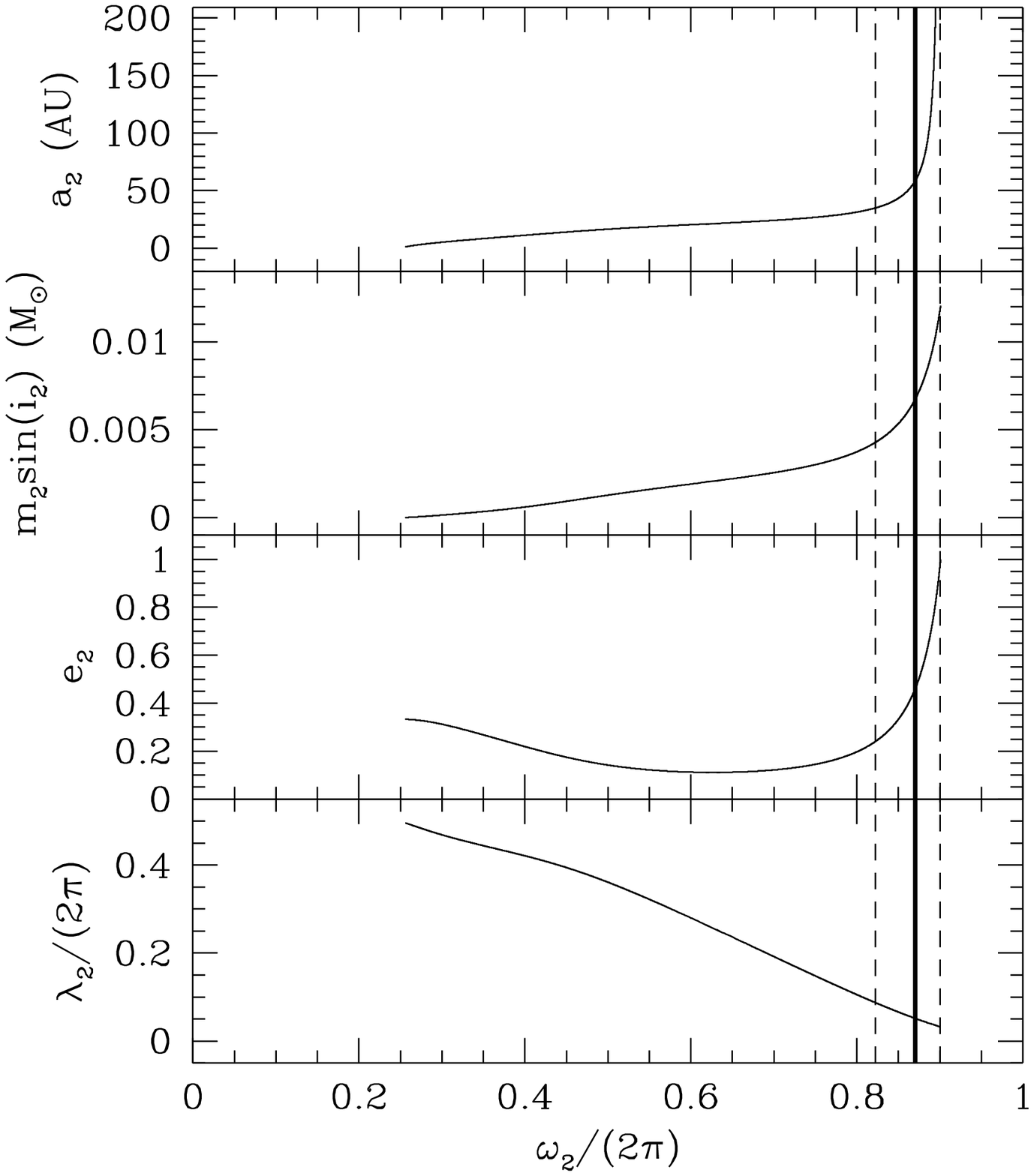}{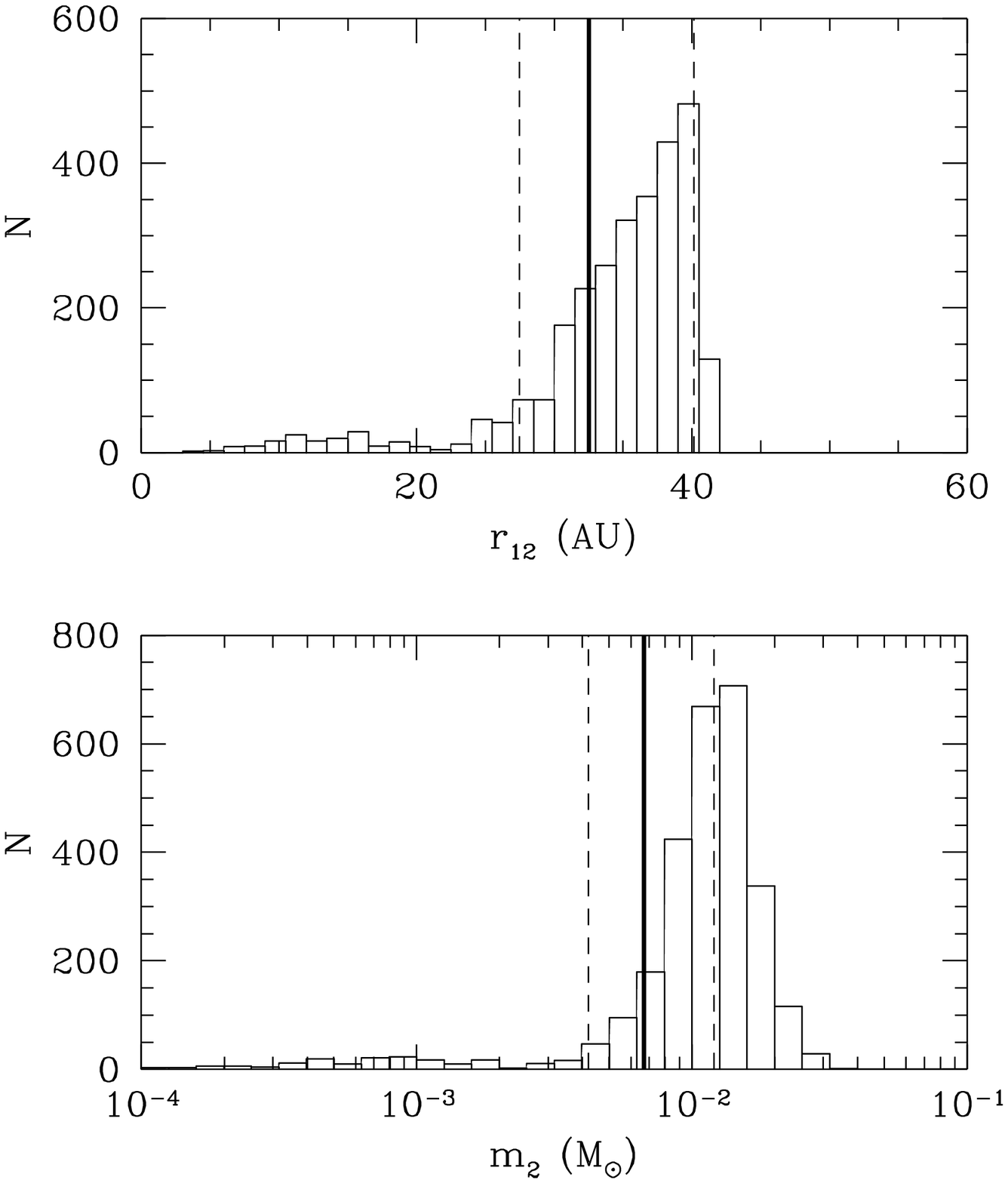}
\caption{\small Allowed values of the semimajor axis $a_2$, mass $m_2$,
eccentricity $e_2$, longitude at epoch $\lambda_2$, and longitude
of pericenter $\omega_2$
for the second companion of PSR~B1620$-$26, using the latest 
available values for four pulse frequency derivatives (left). 
On the right, we show the number of accepted realizations ($N$) 
of the triple for 
different values of $m_2$ and the corresponding distance $r_{12}$
of the second companion from the inner binary in our Monte Carlo 
simulations. Accepted realizations are those leading to short-term
orbital perturbation effects consistent with the current observations.
The vertical solid lines indicates the complete solution obtained by 
including the preliminary value of the fifth derivative. 
The two dashed lines indicate the solutions obtained by decreasing 
or increasing the value of $\fddddd$ by a factor of 1.5.}
\end{figure}

It is extremely reassuring to see that the new 
measurement of $\fddddd$ is consistent with the family of solutions obtained
previously on the basis of the first four derivatives. The implication is that
the signs and magnitudes of these five independently measured quantities are
all consistent with the basic interpretation of the data in terms of a 
second companion orbiting the inner binary pulsar in a Keplerian orbit.
For comparison, the two vertical dashed lines in Figure~1 indicate the
change in the solution obtained by decreasing the value of $\fddddd$ by a
factor of 1.5 (right), or increasing it by a factor of 1.5 (left). Note that 
lower values of $\fddddd$ give higher values for $m_2$. If we
vary the value of $\fddddd$ within its entire $1\sigma$ error bar,
all solutions are allowed, except for the extremely low-mass solutions
with $m_2 \sin i_2 \la 0.002\,M_\odot$. 
In particular, the present $1\sigma$ error on $\fddddd$ does not 
strictly rule out a hyperbolic orbit ($e_2 > 1$) for $m_2$. 
However, it is still possible to derive a strict upper limit on $m_2$ by considering 
hyperbolic solutions and requiring that the relative velocity at infinity of 
the perturber be less than the escape speed from the cluster core.
Indeed, for $m_2 \sin i_2 > 0.055\,M_\odot$, FJRZ find that the 
relative velocity at infinity would exceed the escape speed from the 
cluster core ($\simeq 12\,{\rm km}\,{\rm s}^{-1}$; see Peterson \etal~1995).
A strict  lower limit on the mass,
 $m_2\ga 2\times10^{-4}\,M_\odot$, can also
be set by requiring that the orbital period of the 
second companion be longer than the duration of the timing observations (about
$10\,$yr). Note that
all solutions then give dynamically stable triples, even at the low-mass,
short-period limit (Eggleton \& Kiseleva 1995; JR97).

\subsection{Orbital Perturbations}

Additional constraints and consistency checks on the model can be obtained
by considering the perturbations of the orbital elements of the inner binary
caused by the presence of the second companion. These include a precession of
the pericenter, as well as short-term linear drifts in the inclination and eccentricity.
The drift in inclination can be detected through a change in the
projected semimajor axis of the pulsar. The semimajor axis itself is
not expected to be perturbed significantly by a low-mass second companion
(Sigurdsson 1995).

The latest measurements, obtained by adding a linear drift term to each 
orbital element in the Keplerian fit for the inner binary (TACL) give:
\begin{eqnarray}
\dot \omega_1 & = & -5(8)\times10^{-5}\,{\rm deg}\,{\rm yr}^{-1},\nonumber \\
\dot e_1 & = & 0.2(1.1)\times10^{-15}\,\rm s^{-1}, \nonumber \\
\dot x_p & = & -6.7(3)\times10^{-13}, \nonumber
\end{eqnarray}
where $x_p = a_p \sin i_1$ is the projected semimajor axis of the pulsar. 
Note that only $\dot x_p$ is clearly detected, while the other two measurements 
only provide upper limits.

We have used these measurements in FJRZ to constrain the system by
requiring that all our solutions be consistent with these secular 
perturbations. Specifically, we perform Monte Carlo simulations,
constructing a large number of random realizations of the triple 
system in 3D, and accepting or rejecting them on the basis of
compatibility with the measured orbital perturbations (see JR97
for details). 
The eccentricity of the outer orbit is selected randomly assuming
a thermal distribution, and the other orbital parameters are then
calculated from the standard solutions obtained in \S2.1.
The unknown inclination angles $i_1$ and 
$i_2$ are generated assuming random orientations of the orbital planes, 
and the two position angles of the second companion are determined using
$i_1$, $i_2$, $\omega_2$, $\lambda_2$, and an additional undetermined
angle $\alpha$, which (along with $i_1$ and $i_2$) describes the 
relative orientation of the two orbital planes.
The perturbations are calculated analytically for each realization 
of the system, assuming a fixed position of the second companion.
The perturbation equations are given in Rasio (1994) and JR97.

Figure~1 (right panel) shows the resulting probability distributions for the 
mass $m_2$ and the current separation (at epoch) $r_{12}$
of the second companion.
The Monte Carlo trials were performed using only
our standard solution based on four pulse frequency derivatives, 
since the fifth derivative
is still only marginally detected. The solid lines indicates the
values given by the preliminary measurement of $\fddddd$ and assuming
$\sin i_2=1$.
The most probable value of $m_2 \simeq 0.01\,M_\odot$ is 
consistent with the range of values obtained from the complete
solution using the fifth derivative. 
We have also calculated the probability distribution of $\fddddd$ 
predicted by our Monte Carlo simulations, and found that it is 
consistent with the preliminary measurement, providing yet another
independent self-consistency check on our model.

\section{Secular Eccentricity Perturbations}

We now examine the possibility that the inner binary eccentricity was 
induced by the secular gravitational perturbation of the second companion. 
In the hierarchical three-body problem, analytic expressions for the maximum
induced eccentricity and the period of long-term eccentricity oscillations are
available in certain regimes, depending on the eccentricities and relative inclination
of the orbits.

\subsection{Planetary Regime}

For orbits with small eccentricities and a small relative inclination
($i\la 40^{\circ}$), the classical solution for the long-term secular
evolution of eccentricities and longitudes of pericenters can be written
in terms of an eigenvalue and eigenvector formulation
(e.g., Murray \& Dermott 2000; see Rasio 1995 for simplified 
expressions in various limits). This classical solution is valid to all orders
in the ratio of semimajor axes. The eccentricities oscillate
as angular momentum is transferred between the two orbits.  The
precession of the orbits (libration or circulation)  is coupled to the
eccentricity oscillations, but the relative inclination remains approximately
constant.  In this regime, it can be shown that a stellar-mass second 
companion would be necessary to induce the observed eccentricity in the inner binary
(Rasio 1994; 1995).  Such a large mass for the third body has been ruled out
by recent timing data (see \S 2), implying that secular perturbations
from a third body in a nearly coplanar and circular orbit does not
explain the observed inner eccentricity.  

Sigurdsson (1995) has suggested that it may be possible for the secular
perturbations to grow further because of random distant interactions of the triple
with other cluster stars at distances $\sim100\,$AU, which would perturb
the long-term phase relation between the inner and outer orbits, allowing
the inner eccentricity to ``random walk'' up to a much larger value.
However, the current most probable solution from the timing data requires the
second companion to be in a very wide orbit (separation $\ga40\,$AU), giving
it a very short lifetime in the cluster, and leaving it extremely vulnerable
to disruption by such repeated weak encounters 
(see JR97 for a more detailed discussion).

\subsection{High-Inclination Regime}

For a triple system formed through a dynamical interaction, there is no reason
to assume that the relative inclination of the two orbits should be small.
When the relative inclination of the two orbital planes is $\ga
40^{\circ}$, a different regime of secular perturbations is encountered.
This regime has been studied in the past using the quadrupole
approximation (Kozai 1962; see Holman, Touma, \& Tremaine 1997 for a recent
discussion). Here the relative inclination $i$ of the two
orbits and the inner eccentricity $e_1$ are coupled by the integral
of motion $\Theta = (1-e_1^2) \cos^2 i$ (Kozai's integral). Thus, the
amplitude of the eccentricity oscillations is determined
by the relative inclination. It can be shown that large-amplitude eccentricity
oscillations are possible only when $\Theta < 3/5$ (Holman \etal~1997).
For an initial eccentricity $e_1\simeq0$ and initial
inclination $i_0$ this implies $i_0 > \cos^{-1}\sqrt{3/5}\simeq 40^\circ$
and the maximum eccentricity is then given by
$e_{1\rm max}= [1 - (5/3)\cos^2 i_0 ]^{1/2}$, which approaches
unity for relative inclinations approaching $90^{\circ}$.  
For a sufficiently large relative inclination, this suggests
that it should always be possible to induce an arbitrarily large
eccentricity in the inner binary, and that this could provide an
explanation for the anomalously high eccentricity of the binary
pulsar in the PSR~B1620$-$26 system (Rasio, Ford, \& Kozinsky 1997).
However, there are two additional conditions that must be
satisfied for this explanation to hold.

First, the timescale for reaching a high eccentricity must be
shorter than the lifetime of the system. Although the masses,
initial eccentricities, and ratio of semimajor axes do not affect the maximum
inner eccentricity, they do affect the period of the eccentricity
oscillations (see Holman \etal\ 1997; Mazeh \etal\ 1996).  
The inner longitude of periastron precesses with this period, which can be
quite long, sometimes exceeding the lifetime of the system in the cluster. 
The masses also affect the period, but they decrease the
amplitude of the eccentricity oscillations only when the mass ratio
of the inner binary approaches unity (FKR). 
In Figure~2 we compare the period of the eccentricity oscillations
to the lifetime of the triple in M4. It is clear that for most solutions
the timescale to reach a large eccentricity exceeds the lifetime of the triple.
The only possible exceptions are for very low-mass planets 
($m_2\la 0.002\,M_\odot$) and with the triple
residing far outside the cluster core. These cannot be excluded, 
but are certainly not favored by the observations (see \S 2).

The second problem is that other sources of perturbations may become significant
over these long timescales. In particular, for an inner binary containing compact
objects, general relativistic effects can become important. 
This turns out to play a crucial role for the PSR~B1620$-$26 system,
and we address this question in detail in the next section.

\subsection{General Relativistic Effects}

Additional perturbations that affect the longitude of periastron can
indirectly change the evolution of the eccentricity of the inner
binary in a hierarchical triple.  
For example, tidally- or rotationally-induced quadrupolar 
distortions, as well as general relativity, can cause a significant precession 
of the inner orbit for a sufficiently compact binary.  
If this additional precession is much
slower than the precession due to the secular perturbations, then the
eccentricity oscillations are not significantly affected.  However, if
the additional precession is faster than the secular perturbations,
then eccentricity oscillations are severely damped (see, e.g., Holman \etal\ 
1997).  In addition, if the two precession periods are comparable,
then a type of resonance can occur that leads to a significant increase in the
eccentricity perturbation (FKR).  

Figure~2 compares the various precession periods for PSR~B1620-26 as a
function of the second companion mass $m_2$ for the entire 
one-parameter family of standard solutions constructed in \S 2.
The precession period $P_{{\rm High}~i}$ for high-inclinations was
calculated using the approximate analytic expression given by 
Holman \etal~(1997, eq.~3), while $P_{{\rm Low}~i}$
is from the classical solution (\S 3.1). The general relativistic precession
period for the inner binary, $P_{\rm GR}$, is also shown.
Note that, although we have labeled the plot assuming $\sin i_2=1$, only 
$P_{{\rm Low}~i}$ has an explicit dependence on $m_2$ (and it is calculated
here for $\sin i_2=1$). 

For nearly all solutions we find that the general relativistic precession
is {\it faster\/} than the precession due to the Newtonian secular perturbations.
The only exceptions are for low-mass second companions ($m_2\la
0.005\,M_\odot$) in low-inclination orbits. However, we have already
mentioned above (\S 3.1) that in this case the maximum induced eccentricity
could not reach the present observed value. Most remarkably, however,
we also see from Figure~2 that, for the most probable 
solution (based on the
current measured value of $\fddddd$ and indicated by the vertical solid line 
in the figure), the two precession periods
are {\it nearly equal\/} for a low-inclination system. This suggests that
resonant effects may play an important role in this system, a
possibility that we have explored in detail using numerical integrations.
We describe the numerical results in the next section.

\begin{figure}[t]
\plottwo{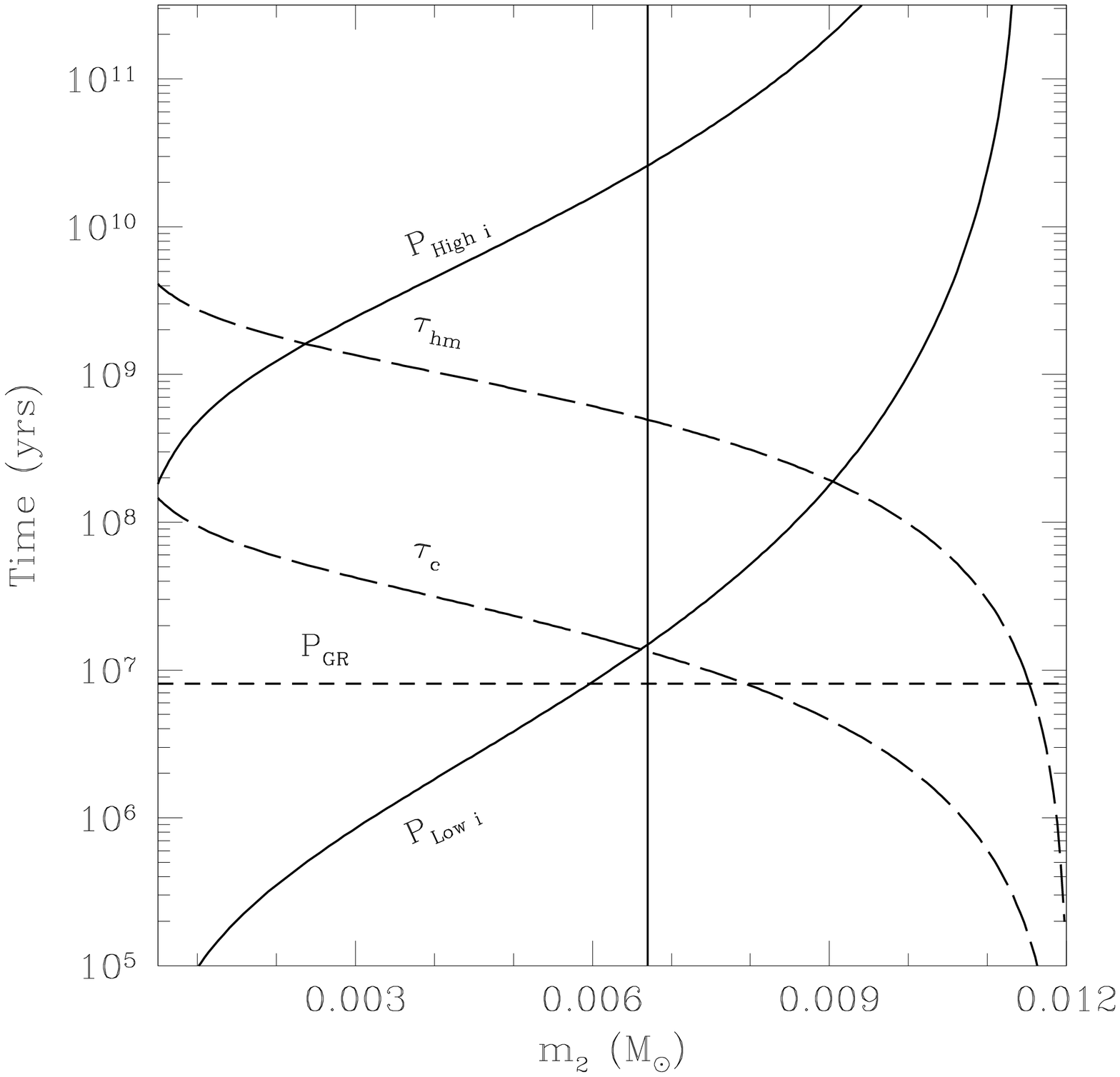}{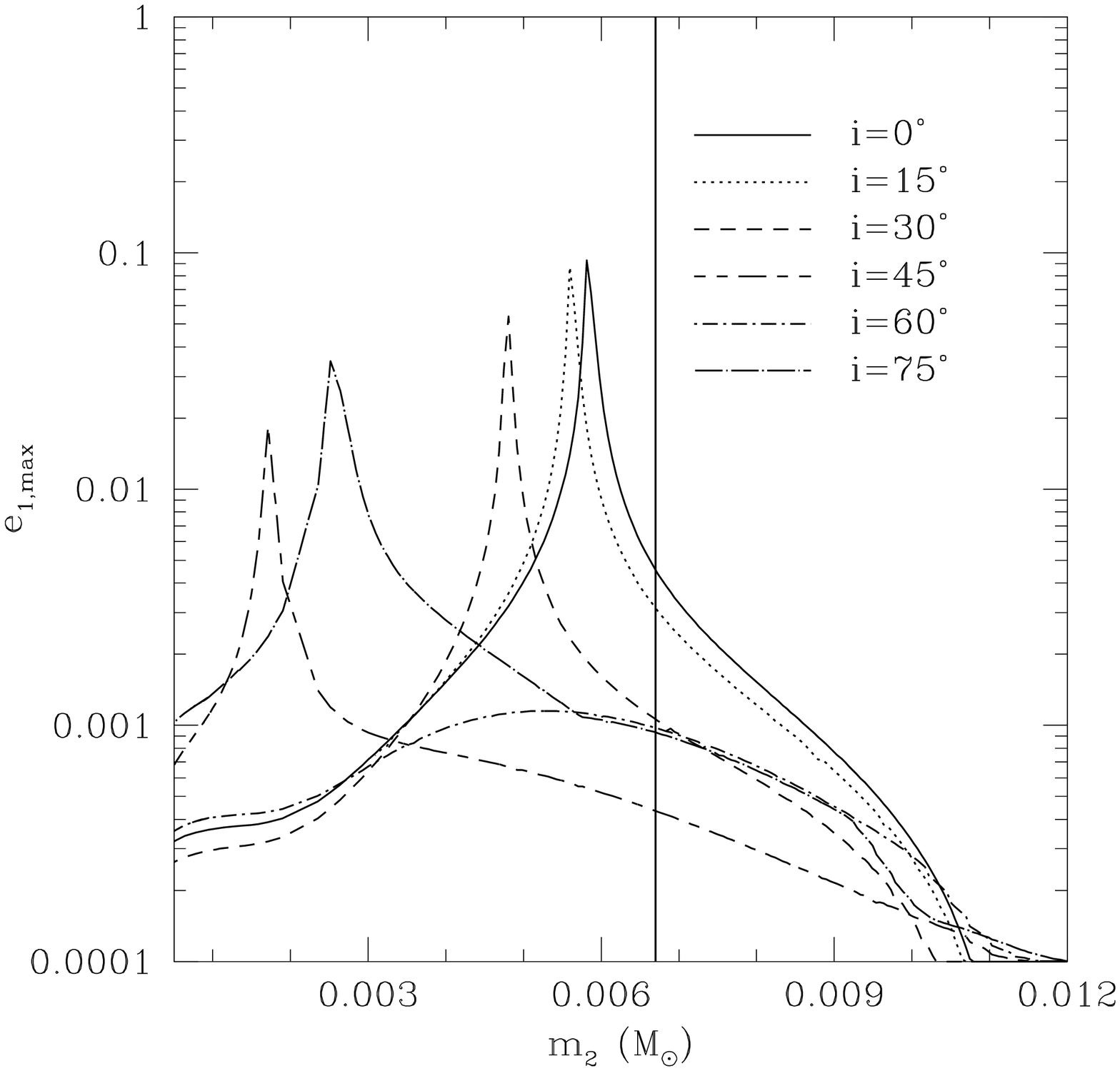}
\caption{\small Comparison of the various secular precession timescales 
in the PSR~B1620$-$26 triple, as a function of the mass of the second 
companion in the standard solution of \S 2 (left). Also shown for comparison 
is the lifetime of the triple, both in the cluster core ($\tau_{\rm c}$) 
and at the half-mass radius ($\tau_{\rm hm}$). 
On the right, we show
the maximum eccentricity of the binary pulsar induced by secular
perturbations in the triple (including the Newtonian perturbation by the
second companion, and the general relativistic precession of the inner
orbit), as a function of the mass of the second companion in the one-parameter
family of solutions from \S 2. Note the clear resonant interaction between
the Newtonian and relativistic perturbations.}
\end{figure}

\subsection{Numerical Integrations of the Secular Evolution Equations}

If the quadrupole approximation used for the high-inclination regime
is extended to octupole order, the resulting secular perturbation equations  
approximate very well the long-term dynamical evolution of hierarchical 
triple systems for a wide range of masses, eccentricities, and inclinations 
(FKR; Krymolowski \& Mazeh 1999).  
We have used the octupole-level secular perturbation
equations derived by FKR to study the long-term eccentricity
evolution of the PSR~B1620$-$26 triple. We integrate
the equations using the variables $e_1 \sin \omega_1$, $e_1 \cos \omega_1$, 
$e_2 \sin \omega_2$, and $e_2 \cos \omega_2$, where $\omega_1$ and $\omega_2$ 
are the longitudes
of pericenters. This avoids numerical problems for nearly circular orbits,
and also allows us to incorporate easily the
first-order post-Newtonian correction into our integrator, which is
based on the Burlish-Stoer method.
We assume that the present inner eccentricity is due entirely to the 
secular perturbations and that the initial eccentricity of the binary pulsar
was much smaller than its present value.
In addition, we restrict our attention to the standard one-parameter
family of solutions constructed in \S 2.1. From the numerical integrations
we can then determine the maximum induced eccentricity.

In Figure~2 (right panel) we show this maximum induced eccentricity in the
inner orbit as a function of the second companion mass for several
inclinations.  For most inclinations and masses, we see that the maximum
induced eccentricity is significantly smaller than
the observed value, as expected from the discussion of \S 3.3.  
However, for a small but significant range of masses near the most
probable value (approximately 
$0.0055\,M_\odot < m_2 < 0.0065\,M_\odot$),
the induced eccentricity for low-inclination systems can reach values
$\ga 0.02$.

As already pointed out in \S 3.2, the maximum induced eccentricity may 
also be limited by the lifetime of the
triple system. From Figure~2 we see that, near resonance, we expect 
$P_{{\rm Low}~i}\simeq P_{\rm GR}\simeq 10^7\,$yr, which is comparable to
the lifetime of the triple in the cluster core, and much shorter than the lifetime
outside of the core. For solutions near a resonance the inner
eccentricity $e_1$ initially grows linearly at approximately the same rate
as it would without the general relativistic perturbation.  However, the
period of the eccentricity oscillation can be many times the period
of the classical eccentricity oscillations.  Although this allows the
eccentricity to grow to a larger value, the timescale for
this growth can then be longer than the expected lifetime of the triple
in the cluster core. For example, with $m_2=0.006\,M_\odot$
we find that the inner binary reaches an eccentricity
of 0.025 after about 1.5 times the expected lifetime of the triple
in the core of M4. Thus, even if the system is near resonance, it
must probably still be residing somewhat outside the core for the secular
eccentricity perturbation to have enough time to grow to the currently
observed value.

\section{Summary}

Our theoretical analysis of the latest timing data for PSR~B1620$-$26
clearly confirms the triple nature of the system. Indeed, the values of
all five measured pulse frequency derivatives are consistent with our basic 
interpretation of a binary pulsar perturbed by the gravitational influence of a
more distant object on a bound Keplerian orbit.
The results of our Monte-Carlo simulations
based on the four well-measured frequency derivatives 
and preliminary measurements of short-term orbital perturbation effects
in the triple are consistent with the complete solution obtained 
when we include the preliminary measurement of the fifth frequency 
derivative. This complete solution corresponds to a 
second companion of mass
$m_2 \sin i_2 \simeq 7\times10^{-3}\,M_\odot$ in an orbit of eccentricity
$e_2\simeq 0.45$ and semimajor axis $a_2\simeq 60\,$AU (orbital period
$P_2\simeq300\,$yr). 
Although the present formal $1\sigma$ error on $\fddddd$ is large, we do not
expect this solution to change significantly as more timing data become
available.

It is possible that the dynamical interaction that formed the triple
also perturbed the eccentricity of the binary pulsar to the anomalously
large value of $0.025$ observed today. However, we have shown that,
through a subtle interaction between the general relativistic corrections
to the binary pulsar's orbit and the Newtonian gravitational perturbation
of the planet, this eccentricity could also have been induced by long-term
secular perturbations in the triple after its formation. The interaction
arises from the near equality between the general relativistic precession
period of the inner orbit and the period of the Newtonian secular
perturbations for a low-inclination system. It allows the eccentricity 
to slowly build up to the presently observed value, on a timescale that
can be comparable to the lifetime of the triple in M4.

All dynamical formation scenarios have to confront the problem that the
lifetimes of both the current triple and its parent star--planet system
are quite short, typically $\sim10^7-10^8\,$yr as they approach
the cluster core, where the interaction is most likely to occur. Therefore,
the detection of a planet in orbit  around the PSR~B1620$-$26 binary
clearly suggests that large numbers of these wide star--planet systems
must exist in globular clusters, since most of them will be destroyed before
(or soon after) entering the core, and most planets will not be able to
survive long in a wide orbit around any millisecond pulsar system
(where they may become detectable through high-precision pulsar timing).
Although a star--planet separation $a_i\sim50\,$AU may seem quite large
when compared to the orbital radii of all recently detected extrasolar planets
(which are all smaller than a few AU; see Marcy \& Butler 1998), one must
remember that the current Doppler searches are most sensitive to planets in 
short-period orbits, and that they could never have detected a low-mass 
companion with an orbital period $\gg 10\,$yr. Similarly, the recent HST search
for planetary transits in 47~Tuc by Gilliland \etal\ (2000), which did not
detect any planet, was only sensitive to very short orbital periods $\la10\,$d.
In addition, it is of course
possible that the parent system may have been a primordial binary star with a 
low-mass, brown dwarf component, rather than a main-sequence star with planets.

\acknowledgements
Many MIT students have contributed significantly to this
work, including most recently J.\ Bostick, E.B.\ Ford, K.J.\ Joshi, 
B.\ Kozinsky, J.\ Madic, and B.\ Zbarsky.
This work was supported by NSF Grant AST-9618116 and NASA ATP Grant
NAG5-8460, and by a Sloan Research Fellowship.

\end{document}